\documentclass[11pt, a4paper]{article}
\pdfoutput=1

\usepackage[height=8.85in,width=6.75in]{geometry}

\usepackage{graphicx,rotating}     
\usepackage[bookmarksopen,colorlinks=true,linkcolor=light_blue,
citecolor=light_pink,urlcolor=dark_red,linktoc=all]{hyperref}

\usepackage{amsmath,tensor}
\usepackage{mathtools}
\usepackage{amsfonts}
\usepackage{amssymb}
\usepackage{slashed}
\usepackage{braket}
\usepackage{bm}
\usepackage{bbm}
\usepackage{cite}
\usepackage{comment}
\usepackage{mathrsfs}
\usepackage{xcolor}
\usepackage[utf8]{inputenc}
\usepackage[footnotesize]{caption}
\usepackage{bbold}
\usepackage{xfrac}
\usepackage{physics}
\usepackage{array}
\usepackage{multirow}

\usepackage[lf]{Baskervaldx} 
\usepackage[bigdelims,vvarbb]{newtxmath} 
\usepackage[cal=boondoxo]{mathalfa} 

\usepackage{chngcntr}
\counterwithin{equation}{section}

\usepackage{multicol}
\definecolor{dark_red}{rgb}{0.7, 0., 0.}
\definecolor{light_pink}{rgb}{1,0.4,0.4}
\definecolor{light_blue}{rgb}{0.284602,0.317763,0.963947}
\definecolor{cred}{RGB}{180,50,40}
\definecolor{darkgreen}{RGB}{0, 100, 0}
\definecolor{desy_blue}{HTML}{009EE2}
\definecolor{desy_orange}{HTML}{FD8800}
\definecolor{forestgreen}{HTML}{228B22}
\definecolor{ochre}{HTML}{CCAA2B}

\newcommand{\Mpl}{M_{\text{Pl}}}

\definecolor{darkred}{rgb}{0.7, 0., 0.}
\definecolor{orangered}{rgb}{1,0.27,0.}
\definecolor{steelblue}{rgb}{0.275,0.51, 0.706}
\definecolor{forestgreen}{rgb}{0.13,0.55,0.13}
\definecolor{brightgreen}{cmyk}{0.75, 0.02, 1.00, 0.00}
\definecolor{sagegreen}{rgb}{0.5, 0.65, 0.4}
\definecolor{sepia}{rgb}{0.55, 0.45, 0.3}

\begin{document}

\hypersetup{pageanchor=false}
\begin{titlepage}

\begin{center}

\hfill KEK-TH-2714\\
\hfill RESCEU-7/25\\
\hfill IPMU25-0017 \\
\hfill CTPU-PTC-25-14

\vskip .6in

{\Huge \bfseries
Increase of \boldsymbol{$n_s$} in regularized pole inflation\\
\& Einstein--Cartan gravity\\
}
\vskip .8in

{\Large Minxi He$^{a}$, Muzi Hong$^{b,c,d}$, Kyohei Mukaida$^{e,f}$}

\vskip .3in
\begin{tabular}{ll}
  $^a$& \!\!\!\!\!\emph{Particle Theory and Cosmology Group, Center for Theoretical Physics of the Universe, }\\[-.3em]
  & \!\!\!\!\!\emph{Institute for Basic Science (IBS),  Daejeon, 34126, Korea}\\
  $^b$& \!\!\!\!\!\emph{Department of Physics, Graduate School of Science, The University of Tokyo, Tokyo 113-0033, Japan}\\
  $^c$& \!\!\!\!\!\emph{RESCEU, Graduate School of Science, The University of Tokyo, Tokyo 113-0033, Japan}\\ 
  $^d$& \!\!\!\!\!\emph{Kavli IPMU (WPI), UTIAS, The University of Tokyo, Kashiwa 277-8583, Japan}\\
  $^e$& \!\!\!\!\!\emph{Theory Center, IPNS, KEK, 1-1 Oho, Tsukuba, Ibaraki 305-0801, Japan}\\
  $^f$& \!\!\!\!\!\emph{Graduate University for Advanced Studies (Sokendai), 1-1 Oho, Tsukuba, Ibaraki 305-0801, Japan}
  \end{tabular}

\end{center}
\vskip .6in

\begin{abstract}
\noindent
We show that the regularization of the second order pole in the pole inflation can induce the increase of $ n_s $, which may be important after the latest data release of cosmic microwave background (CMB) observation by Atacama Cosmology Telescope (ACT). 
Pole inflation is known to provide a unified description of attractor models that they can generate a flat plateau for inflation given a general potential.
Recent ACT observation suggests that the constraint on the scalar spectral index $ n_s $ at CMB scale may be shifted to a larger value than the predictions in the Starobinsky model, the Higgs inflation, and the $\alpha$-attractor model, which motivates us to consider the modification of the pole inflation. 
We find that if we regularize the second order pole in the kinetic term such that the kinetic term becomes regular for all field range, we can generally increase $ n_s $ because the potential in the large field regime will be lifted.
We have explicitly demonstrated that this type of regularized pole inflation can naturally arise from the Einstein--Cartan formalism, and the inflationary predictions are consistent with the latest ACT data without spoiling the success of the $\alpha$-attractor models.
\end{abstract}

\end{titlepage}


\section{Introduction}
\label{sec:intro}

The cosmic inflation~\cite{Starobinsky:1980te,Sato:1980yn,Guth:1980zm,Mukhanov:1981xt,Linde:1981mu,Albrecht:1982wi}\footnote{See Ref.~\cite{Sato:2015dga} for a review.} is a leading paradigm describing the extremely early stage of the Universe.
It solves the flatness/horizon problems and moreover provides a mechanism for the generation of primordial density perturbations by the accelerated expansion of the Universe.
Such an exponential expansion phase is driven by a scalar field, called inflaton, which is slowly rolling down its potential.
The recent observations of cosmic microwave background (CMB) implies a concave potential for the inflaton~\cite{BICEP:2021xfz}.

The Starobinsky inflation~\cite{Starobinsky:1980te} and the Higgs inflation~\cite{Cervantes-Cota:1995ehs,Bezrukov:2007ep,Barvinsky:2008ia} have drawn much attention as representative examples of inflation not only for their simplicity but also for their successful predictions of the scalar spectral index $n_s$ and the tensor-to-scalar ratio $r$~\cite{BICEP:2021xfz}.
The inflationary predictions of these models are essentially the same where $n_s = 1 - 2/ N_e $
and $r = 12/ N_e^2$ with $N_e$ being the e-folding number of inflation.
This coincidence of predictions has driven extensive studies on the attractor of inflation models, where the broad classes of seemingly different models lead to the same predictions~\cite{Kallosh:2013maa,Kallosh:2013tua,Giudice:2014toa,Pallis:2014dma,Kallosh:2014laa,Mosk:2014cba,Pieroni:2015cma}. 
Lately, the first class of the attractor including the Starobinsky inflation is generalized to the so-called $\alpha$-attractor models~\cite{Ellis:2013nxa,Ferrara:2013rsa,Kallosh:2013yoa,Kallosh:2014rga,Carrasco:2015pla,Roest:2015qya,Linde:2015uga,Scalisi:2015qga}, whose prediction is now shifted to $n_s = 1 - 2/ N_e $
and $r = 12 \alpha / N_e^2$ with $\alpha$ being a new parameter characterizing this attractor.
Notable feature of these scenarios is insensitivity on the details of the inflaton potential, and this is the reason why they are called the attractor models.

These attractors can be understood in a unified framework of the so-called pole inflation~\cite{Galante:2014ifa,Broy:2015qna,Terada:2016nqg,Fu:2022ypp,Aoki:2022bvj,Karamitsos:2021mtb,Pallis:2021lwk,Dias:2018pgj,Saikawa:2017wkg,Kobayashi:2017qhk}.\footnote{It is also related to the running kinetic inflation~\cite{Takahashi:2010ky,Nakayama:2010kt}.} There the flatness of the potential is characterized by an enhancement of the kinetic term of the inflaton, which is controlled by the order of a pole in the field space.
Indeed all the models mentioned above have the second order pole in the kinetic term in the Einstein frame, $- \gamma^2 \Mpl^2 \phi^{-2} (\partial \phi)^2 / 2$. The inflationary prediction is given by $n_s = 1 - 2/ N_e $
and $r = 8 \gamma^2 / N_e^2$, which is controlled by the residue of the pole $\gamma^2$.
As long as the location of the pole in the field space is different from the potential minimum, the potential becomes flattened as the inflaton field approaches the pole after the canonical normalization of the inflaton field, and the inflaton potential asymptotically approaches a non-vanishing constant value in the limit of $\phi \to \pm 0$.

Recently, the Atacama Cosmology Telescope (ACT) has released the latest data of CMB observation~\cite{ACT:2025fju,ACT:2025tim}, which shows a slight increase of the previous result on $n_s$ (see also \cite{Dioguardi:2025mpp,Kim:2025dyi,Antoniadis:2025pfa,Salvio:2025izr,Gialamas:2025kef,Dioguardi:2025vci,Berera:2025vsu,Aoki:2025wld,Kallosh:2025rni,Gao:2025onc,Brahma:2025dio}).
In particular, the attractor inflations mentioned above, such as the Starobinsky inflation, the Higgs inflation, and the $\alpha$-attractor models, are now disfavored at $2\sigma$ level.
The main purpose of this paper is to point out that the regularization of the pole in the kinetic term can lead to an increase of $n_s$ and thereby explain the new ACT data without spoiling the success of these attractor models, where $- \gamma^2 (\phi^{2}/\Mpl^2 + \lambda^2)^{-1} (\partial \phi)^2 / 2$.
We dub this new model as a \textit{regularized pole inflation}.
We also show that the regularization of the pole is naturally realized in the Einstein--Cartan (EC) gravity, where the scalaron can be identified with the inflaton.

The paper is organized as follows. 
In Sec.~\ref{sec:regulatedpole}, we discuss the general properties of the regularized pole inflation, deriving the inflation predictions and identify the deviations from the pole inflation models. 
In Sec.~\ref{sec:ECregulated}, we show one realization of regularized pole inflation in the framework of EC gravity as a typical example. 
Also, we show the predictions on the $ n_s-r $ plot given the latest results from ACT. 
Finally, we conclude and summary our results in Sec.~\ref{sec:conclusion}.

\section{Regularized pole inflation}\label{sec:regulatedpole}

Let us start with a single-field model as  
\begin{align}
    S= \int \dd^4x ~ \sqrt{-g} \left[ \frac{\Mpl^2}{2} R -\frac{1}{2} K(\phi) g^{\mu\nu} \partial_{\mu} \phi \partial_{\nu} \phi -V (\phi) \right] ~,
\end{align}
where the potential $V(\phi)$ is assumed to be smooth in the relevant field space of $\phi$,
while $ K(\phi) $ is an arbitrary function of $ \phi $ which can come from, for example, Weyl transformation (or conformal transformation)~\cite{Maeda:1987xf,Maeda:1988ab} when removing the non-minimal coupling between $ R $ and $ \phi $. $ K(\phi)=1 $ corresponds to the canonical case. 
Here, we consider the pole inflation scenario~\cite{Galante:2014ifa,Broy:2015qna,Terada:2016nqg} where $ K(\phi) $ generally takes the form by focusing on the highest order pole as 
\begin{align}
    K(\phi) = \frac{\gamma^2 \Mpl^p}{\left( \phi - \phi_0 \right)^p} ~,
\end{align}
where $ \gamma > 0 $ and $ \phi_0 $ are constants and $ p $ is a non-negative integer characterizing the order of the pole of $ K(\phi) $. 
In particular, $ p=2 $ is also known as the $ \alpha $-attractor model~\cite{Ellis:2013nxa,Ferrara:2013rsa,Kallosh:2013yoa,Kallosh:2014rga,Carrasco:2015pla,Roest:2015qya,Linde:2015uga,Scalisi:2015qga}. 
We can always shift the field value $ \phi \to \phi+\phi_0 $ such that the pole is located at the origin 
\begin{align}
    S= \int \dd^4x ~ \sqrt{-g} \left[ \frac{\Mpl^2}{2} R -\frac{1}{2} \frac{\gamma^2 \Mpl^p}{\phi^p} g^{\mu\nu} \partial_{\mu} \phi \partial_{\nu} \phi -V (\phi+\phi_0) \right] ~. 
\end{align}
In this form, we can see that the positive and negative $ \phi $ branchs are separated by the pole at the origin of the real axis because the kinetic term diverges at $ \phi=0 $. 
Note that for $ p $ to be odd number, the kinetic term sign will flip if one takes $ \phi <0 $, which causes ghost instability. 
In the following, we consider even $ p $ to avoid such a case. 

Canonicalizing the kinetic term by 
\begin{align}
    \varphi =\pm  \int\frac{\gamma \Mpl^{p/2}}{\phi^{p/2}} \dd\phi ~, 
\end{align}
where $ \pm $ represents two independent solutions in each branch, one can transform the model into a canonical single-field model. 
Specifically, there are essential difference for $ p \neq 2 $ and $ p=2 $~\cite{Galante:2014ifa,Broy:2015qna,Terada:2016nqg}. 
We focus on the $ p=2 $ case in this paper, and hence the canonicalization gives 
\begin{align}\label{eq-separate-branch}
    \frac{\phi}{\Mpl} = {\rm sign} (\phi) \exp \left( \pm \frac{\varphi}{\gamma \Mpl} \right) ~.
\end{align}
The potential can be expanded around the pole as
\begin{align}
    V (\phi) \simeq V_0 + V_1 \phi + \cdots ~, \qquad V_0 \equiv V(\phi_0) ~, \quad V_1 \equiv V'(\phi_0)~.
\end{align}
Here we assume that the potential energy is positive at the location of the pole $V_0 > 0$ to have a successful inflation.
We obtain a potential as a combination of exponential functions of $ \phi $ 
\begin{align}
    V(\phi(\varphi)) = V_0 +  V_1 \Mpl \operatorname{sign} (\phi) \exp \left( \pm \frac{\varphi}{\gamma \Mpl} \right)  + \cdots ~.
\end{align}
In this case, $ \phi =0 $ is pushed to $ \varphi = \mp \infty $ by the field redefinition so $ \phi $ can never reach the pole. 
Near the pole, the kinetic term is largely enhanced, or equivalently, the potential becomes flat where inflation can occur and give a nearly scale-independent power spectrum of curvature perturbation at CMB scale, \textit{i.e.} $ n_s \simeq 1 $.
As long as the system resides close enough to the attractor, the higher order terms in $\phi$ are negligible.
This is the reason why this model is regarded as an attractor model where the predictions are insensitive to the details of the potential.
The inflationary predictions in this case are given by
\begin{equation}
    n_s = 1 - \frac{2}{N_e} ~, \qquad r = \frac{8 \gamma^2}{N_e^2} ~,
\end{equation}
which reproduces the result of the Starobinsky inflation for $\gamma^2 = 3/2$ for instance.

\begin{figure}[t]
    \centering
    \includegraphics[width=0.45\linewidth]{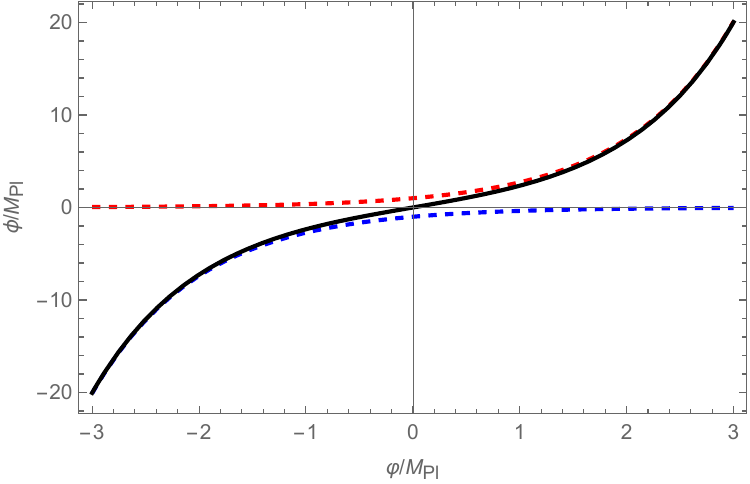}
    \caption{The red and blue dashed lines correspond to $ {\rm sign} (\phi) >0 $ with $ + $ and $ {\rm sign} (\phi) <0 $ with $ - $ in solutions~\eqref{eq-separate-branch}, respectively. 
    The black line corresponds to the solutions~\eqref{eq-connect-branch} with $ + $ sign.}
    \label{fig-connection}
\end{figure}

Now, we consider a new type of inflation model based on the above discussion. 
We take $ p=2 $ as an example.
We regularize the pole by adding an additional constant $ \lambda \in \mathbb{R} $ (we simply take $ \lambda >0 $ without loss of generality) in the denominator of the shifted $ K(\phi) $ such as
\begin{align}
    K_{\rm re}(\phi) = \frac{\gamma^2}{ \phi^2/\Mpl^2 + \lambda^2} ~.
\end{align}
After regularization, there is no pole on the real axis but there are two on the imaginary axis, namely $ \pm i \,\lambda\, \Mpl $, so we will not hit the pole and the kinetic term will not diverge given $ \phi $ a real scalar field. 
Now $ K_{\rm re} (\phi) $ has a Breit-Wigner type of shape where the width is controlled by $ 2\lambda $ and the height by $ \gamma^2/\lambda^2 $, so the kinetic term will still be enhanced when $ \phi $ crosses the origin which is now a regular point. 
Such a kinetic term can also be canonicalized with 
\begin{align}\label{eq-connect-branch}
    \frac{\phi}{\Mpl} = \pm \lambda\sinh \frac{\varphi}{\gamma \Mpl} ~. 
\end{align}
It is interesting to see that this solution is connecting the two separated branches before regularization, as the hyperbolic sine function is the combination of two exponential functions. 
For example, the above solution with $ + $ sign is connecting the $ + $ solution in $ {\rm sign} (\phi) >0 $ branch and $ - $ solution in $ {\rm sign} (\phi) <0 $ branch in Eq.~\eqref{eq-separate-branch}, as shown in Fig.~\ref{fig-connection}. 

\begin{figure}[t]
    \centering
    \vskip-1em
    \includegraphics[width=0.9\linewidth]{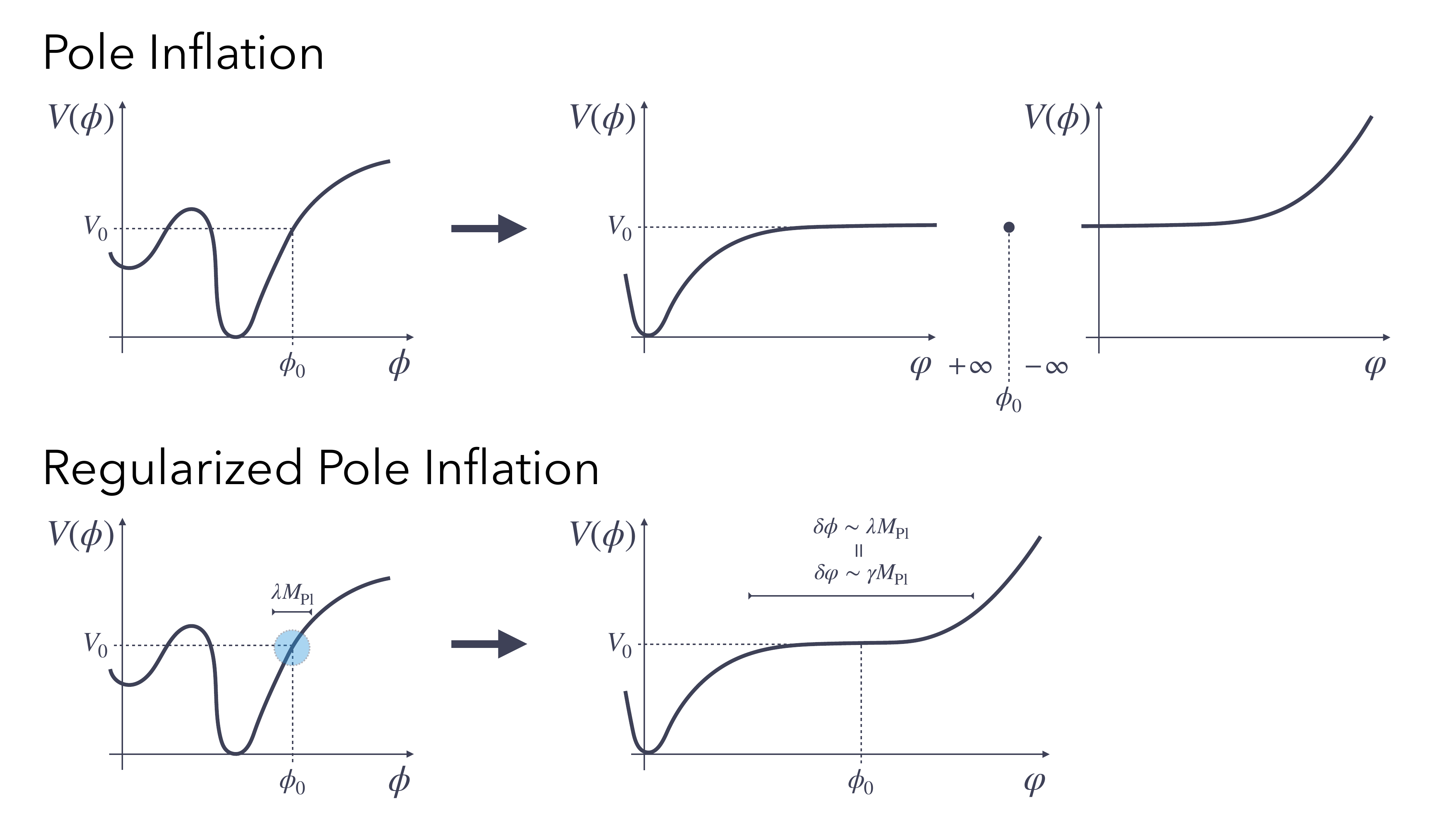}
    \caption{Schematic figure describing the idea of the regularized pole inflation.
    In the pole inflation, the potential at the pole $\phi_0$ is exponentially flattened by the divergence of the kinetic term, which results in two separated branches.
    On the other hand, in the regularized pole inflation, the kinetic term is enhanced within a finite field range of $|\phi - \phi_0 | < \lambda \Mpl$ while the point $\phi_0$ is regularized.
    As a result, the two branches are smoothly connected. 
    Moreover, since a finite field range of the potential is probed by the regularized pole, the attractor behavior is expected only for $\lambda \Mpl \ll \Delta \phi$ with $\Delta \phi$ being the field range where the potential is well approximated by a linear function.
}
    \label{fig-schematic}
\end{figure}

Inserting Eq.~\eqref{eq-connect-branch} into the potential, one readily finds
\begin{align}
    \label{eq-V2-potential}
    V(\phi(\varphi)) \simeq V_0 \pm V_1 \Mpl \lambda \sinh \left( \frac{\varphi -\varphi_0}{\gamma \Mpl} \right) + \cdots ~,
\end{align}
where $ \varphi_0 =\gamma \Mpl {\rm arcsinh} [V_0/(\lambda \Mpl V_1)] $ such that the minimum of the potential is at the origin for convenience.
This potential has an essential difference from the $\alpha$-attractor case.
Specifically, $ V(\varphi) $ grows to infinity for $ |\varphi| \to \infty $ while the potential in the $\alpha$-attractor inflation grows only on one side, which means that the plateau for inflation can be infinitely long in the latter case but finite in the former.
The length of the plateau is controlled by $ \lambda \Mpl V_1/V_0 $ and the Starobinsky limit corresponds to $ \lambda \Mpl V_1/V_0 \to 0 $.
Note here that the negligence of the higher order terms in $\phi$ is not justified in general.
Suppose that the full potential around the pole $\phi_0$ can be well approximated by a linear potential $V(\phi) \simeq V_0 + V_1 \phi$ for a certain field range $|\phi | \lesssim \Delta \phi$.
For the regularized pole inflation, the enhancement of the kinetic term is expected for a finite field range $|\phi |/\Mpl \lesssim \lambda$.
Hence, for $\lambda > \Delta \phi /\Mpl $, the higher order terms in $\phi$ cannot be neglected, which is now sensitive to the details of the potential and thereby cannot be regarded as an attractor.
This consideration puts an upper bound as $\lambda \ll \Delta\phi /\Mpl $.
The idea of the regularized pole inflation is illustrated in Fig.~\ref{fig-schematic}. 
As we will see in the following, such a minor modification of $\lambda$ is sufficient to explain the recent ACT data.

This type of regularization of the pole in the kinetic term can be easily achieved when considering EC framework, as we will see in Sec.~\ref{sec:ECregulated}. 
If we consider the latest result from ACT, the increase of the potential in the large field regime can be important to explain the data.
Without loss of generality, we take $ V_1>0 $, and the solution with plus sign in Eq.~\eqref{eq-V2-potential}. 
In this case, inflation occurs in the $ \varphi>0 $ regime, so we can focus on that during inflation. 
Inflation occurs when both $ \epsilon_V \ll 1 $ and $ |\eta_V| \ll 1 $ are satisfied, and ends at $ \varphi_f= {\rm max} \{ \varphi(\epsilon_V=1), \varphi(|\eta_V|=1)\} $ around $ \varphi =+0 $. 
The purpose of this paper is to see how $ \lambda $ create the deviation from the second order pole inflation, so let us consider the small $ \lambda $ regime to avoid unnecessary complication. 
As a result, we have 
\begin{align}
    \epsilon_V &\simeq \frac{1}{2\gamma^2} \left( 1- e^{\frac{\varphi}{ \gamma \Mpl}} \right)^{-2} + \frac{1}{4\gamma^2} \left( 1- e^{-\frac{\varphi}{\gamma \Mpl}} \right)^{-1} \frac{V_1^2}{V_0^2} \Mpl^2 \lambda^2 + \mathcal{O} \left( \lambda^4 \right) ~, \\
    \eta_V &\simeq \frac{1}{\gamma^2} \left( 1- e^{\frac{\varphi}{ \gamma \Mpl}} \right)^{-1} + \frac{1}{4\gamma^2} \frac{ 1+ e^{\frac{\varphi}{ \gamma \Mpl}} }{1- e^{-\frac{\varphi}{ \gamma \Mpl}}} \frac{V_1^2}{V_0^2} \Mpl^2 \lambda^2 + \mathcal{O} \left( \lambda^4 \right) ~. 
\end{align}
In $ \lambda=0 $ limit, one can see that $ \epsilon_V=1 $ determines the end of inflation when $ \gamma \geq \sqrt{2} $, and $ \eta_V =-1 $ does when $ \gamma < \sqrt{2} $. 
The e-fold number of canonical single-field slow-roll inflation is then calculated as 
\begin{align}
    N_e \simeq \int^{\varphi_*}_{\varphi_f} \frac{1}{\sqrt{2\epsilon_V}} \frac{\dd \varphi}{\Mpl} ~, 
\end{align}
which allows us to express the field value corresponding to the pivot scale $ \varphi_* $ as a function of $ N_e $. 
As a result, the scalar spectral index and the tensor-to-scalar ratio evaluated at pivot scale $ k_* $ up to $ \mathcal{O} \left( \lambda^2 \right) $ and leading order in $ N_e $ are given by 
\begin{align}
    n_s -1 &\simeq -6 \epsilon_V +2 \eta_V 
    \simeq -\frac{2}{N_e} + \frac{2N_e}{3\gamma^4} \frac{V_1^2}{V_0^2} \Mpl^2 \lambda^2 ~, \label{eq-approx-ns} \\
    r &\simeq 16 \epsilon_V 
    \simeq \frac{8\gamma^2}{N_e^2} + \frac{8}{3\gamma^2} \frac{V_1^2}{V_0^2} \Mpl^2 \lambda^2 ~. \label{eq-approx-r} 
\end{align}
These results show that the leading order contributions are the same as the $ \alpha $-attractor model with proper choice of $ \gamma $, and both $ n_s $ and $ r $ increase as the pole regularized by $ \lambda $ and the correction is proportional to $ \lambda^2 $. 

Tu sum up, we have seen the general feature of the regularized pole inflation, in particular, how the predictions on $ n_s $ and $ r $ are shifted by the finite width of the pole.
Next, we will show a concrete example of the regularized pole inflation in EC gravity. 

\section{Realization of regularized pole inflation in Einstein--Cartan gravity}\label{sec:ECregulated}

One realization of regularized pole inflation is inflation in the Einstein--Cartan gravity\footnote{See Ref.~\cite{Trautman:2006fp} for an introduction.}\footnote{ A particular type of realization of regularized pole inflation can also be found in the general metric-affine gravity with an additional scalar field~\cite{Racioppi:2024zva}.}. 
Einstein--Cartan gravity allows non-zero torsion compared with general relativity (GR), while still requiring metricity condition.
One can add torsion components in the Lagrangian, besides the Einstein--Hilbert term, \textit{i.e.}, the Ricci scalar term in GR \cite{Karananas:2021zkl,Salvio:2022suk, He:2024wqv, He:unpublished}.
Here, we consider the Lagrangian containing dimension 4 operators only as a completed square.
It has been shown that this form can lead to a canonically normalizable scalaron, instead of $P(\phi,X)$-type inflation \cite{He:2024wqv}.
Further, we restrict ourselves to the class of Lagrangian where the Ricci scalar only appear as a linear term, such that a Weyl transformation is not necessary.
A general Lagrangian up to dimension 4 satisfying the conditions above is given as  
\begin{align}
    \label{eq-general1}
    \begin{split}
        S = \int \sqrt{-g} \dd^4 x\, & \left[
            \frac{\Mpl^2}{2} \qty(R + \beta_1 S_{\mu} S^{\mu} + \beta_2 T_{\mu} T^{\mu} + \beta_3 S_{\mu} T^{\mu} )
            \right.\\
            & 
            \left.
            + \alpha_\text{R} \qty( \alpha_1 S_{\mu} S^{\mu} + \alpha_2 T_{\mu} T^{\mu} + \alpha_3 S_{\mu} T^{\mu} + \alpha_4 \nabla_\mu S^\mu + \alpha_5 \nabla_\mu T^\mu) ^ 2
            \right]~,
    \end{split}
\end{align}
where $ \nabla_{\mu} $ is the covariant derivative associated with the Levi-Civita connection, $T^\mu$ is the vector component, and $S^\mu$ is the axial vector component of torsion\footnote{It is possible to eliminate either $\alpha_3$ or $\beta_3$ by performing a field redefinition. Here we retain the redundancy for later discussion.}. 
By introducing an auxiliary field $ \chi $ and Legendre transformation, the action becomes 
\begin{align}
    S = \int \sqrt{-g} \dd^4 x\, & \left[
            \frac{\Mpl^2}{2} \qty(R + \beta_1 S_{\mu} S^{\mu} + \beta_2 T_{\mu} T^{\mu} + \beta_3 S_{\mu} T^{\mu} )
            \right. \nonumber \\
            & 
            \left.
            + 2\chi \qty( \alpha_1 S_{\mu} S^{\mu} + \alpha_2 T_{\mu} T^{\mu} + \alpha_3 S_{\mu} T^{\mu} + \alpha_4 \nabla_\mu S^\mu + \alpha_5 \nabla_\mu T^\mu) - \frac{\chi^2}{\alpha_R}
            \right]~,
\end{align}
from which we can eliminate the components of torsion $ S^{\mu} $ and $ T^{\mu} $ by solving the constraints and insert them back to the action~\cite{He:2024wqv}. 
As a result, we have 
\begin{align}
    S = \int \sqrt{-g} \dd^4 x\, & \left[ 
    \frac{\Mpl^2}{2} R -\frac{\chi ^ 2}{\alpha_\text{R}}
    - \frac{1}{2} K (\chi)
    \partial_\mu \chi \partial^\mu \chi    \right] ~,
\end{align}
where the kinetic function is
\begin{align}
    K (\chi) = \frac{8 ( \alpha_2 \alpha_4^2 - \alpha_3 \alpha_4 \alpha_5 + \alpha_1 \alpha_5^2 ) \chi + 2\Mpl^2 ( \beta_2 \alpha_4^2  - \beta_3 \alpha_4 \alpha_5 + \beta_1 \alpha_5^2 )}
    {2 (4 \alpha_1 \alpha_2 - \alpha_3^2) \chi^2 + \Mpl^2 ( 2\alpha_1 \beta_2 - \alpha_3 \beta_3 + 2\alpha_2 \beta_1) \chi + \frac{\Mpl^4}{8} (4 \beta_1 \beta_2 - \beta_3^2)}~.
\end{align}
Now let us consider for example the special cases when
\begin{equation}
    \alpha_2 \alpha_4^2 - \alpha_3 \alpha_4 \alpha_5 + \alpha_1 \alpha_5^2=0~,\quad
    4 \alpha_1 \alpha_2 - \alpha_3^2 < 0~.
\end{equation}
To have a regularized pole inflation we also require
\begin{equation}
    \beta_2 \alpha_4^2  - \beta_3 \alpha_4 \alpha_5 + \beta_1 \alpha_5^2 < 0, \quad
    ( 2\alpha_1 \beta_2 - \alpha_3 \beta_3 + 2\alpha_2 \beta_1)^2 - (4 \alpha_1 \alpha_2 - \alpha_3^2) (4 \beta_1 \beta_2 - \beta_3^2) < 0.
\end{equation}
Thus one can match the results discussed in the last section.

A specific set of parameters is that in Ref.~\cite{He:2024wqv}.
The Ricci scalar $\bar{R}$ in the Einstein--Cartan gravity can be expressed by the Ricci scalar $R$ in general relativity and the torsion components as\footnote{We follow the conventions in Ref.~\cite{He:2024wqv}.} 
\begin{equation}
  \bar{R}  = R + 2 \nabla_{\mu} T^{\mu} - \frac{2}{3} T_{\mu} T^{\mu} + \frac{1}{24} S_{\mu} S^{\mu} + \frac{1}{2} q^{\mu\nu\rho} q_{\mu\nu\rho} ~,
\end{equation}
where $q^{\mu\nu\rho}$ is the tensor component of torsion, which always constrains itself to zero, and thus we drop it from now on.
Using this Ricci scalar in the Einstein--Cartan gravity, the Nieh-Yan term~\cite{Nieh:1981ww,Nieh:2008btw} and the Holst term~\cite{Hojman:1980kv,Nelson:1980ph,Castellani:1991et,Holst:1995pc}, we can construct \cite{He:2024wqv}\footnote{The case of $\alpha_3 = -2/3$ is discussed in \cite{Salvio:2022suk, Salvio:2025izr}. See also \cite{Racioppi:2024zva,Racioppi:2024pno,Gialamas:2024uar} for related scenarios using the Holst term.}
\begin{equation}
S = \int \sqrt{-g} \dd^4 x\, \qty[
    \frac{\Mpl^2}{2}
    \qty( R - \frac{2}{3} T_{\mu} T^{\mu} + \frac{1}{24} S_{\mu} S^{\mu} + \beta_3 S_\mu T^\mu )
    + \alpha_\text{R} \qty(\nabla_\mu S^\mu + \alpha_3 S_\mu T^\mu)^2
  ]~,
\end{equation}
corresponding to $\beta_1 = 1/24$, $\beta_2 = - 2/3$, $\alpha_1=\alpha_2 = \alpha_5 = 0$, and $\alpha_4 = 1$.
The action can thus be written as
\begin{equation}\label{eq-model}
    S = \int \sqrt{-g} \dd^4 x\, \qty[ 
    \frac{\Mpl^2}{2} R 
    - \frac{96 \Mpl^2}{9\qty(4 \alpha_3 \chi + \beta_3 \Mpl^2)^2 + \Mpl^4}
    \frac{\partial_\mu \chi \partial^\mu \chi}{2}
    -\frac{\chi ^ 2}{\alpha_\text{R}}
    ]~,
\end{equation}
which is an example of the regularized pole inflation. 
Following the procedure in Sec.~\ref{sec:regulatedpole},  we can obtain a canonical single-field model as 
\begin{align}
    S = \int \sqrt{-g} \dd^4 x\, \qty[ 
    \frac{\Mpl^2}{2} R 
    - 
    \frac{\partial_\mu \phi \partial^\mu \phi}{2}
    -\frac{\Mpl^4}{144 \alpha_\text{R} \alpha_3^2} \left[ 3\beta_3 +\sinh \left( \sqrt{\frac{3}{2}}\alpha_3 \frac{\phi}{\Mpl} \right) \right]^2
    ]~,
\end{align}
by redefining 
\begin{align}
    \frac{\chi}{\Mpl} = -\frac{\beta_3}{4\alpha_3} \Mpl - \frac{\Mpl}{12\alpha_3} \sinh \left( \sqrt{\frac{3}{2}} \alpha_3 \frac{\phi}{\Mpl} \right) ~.
\end{align}
\begin{figure}[t]
  \centering
  \includegraphics[width=0.5\linewidth]{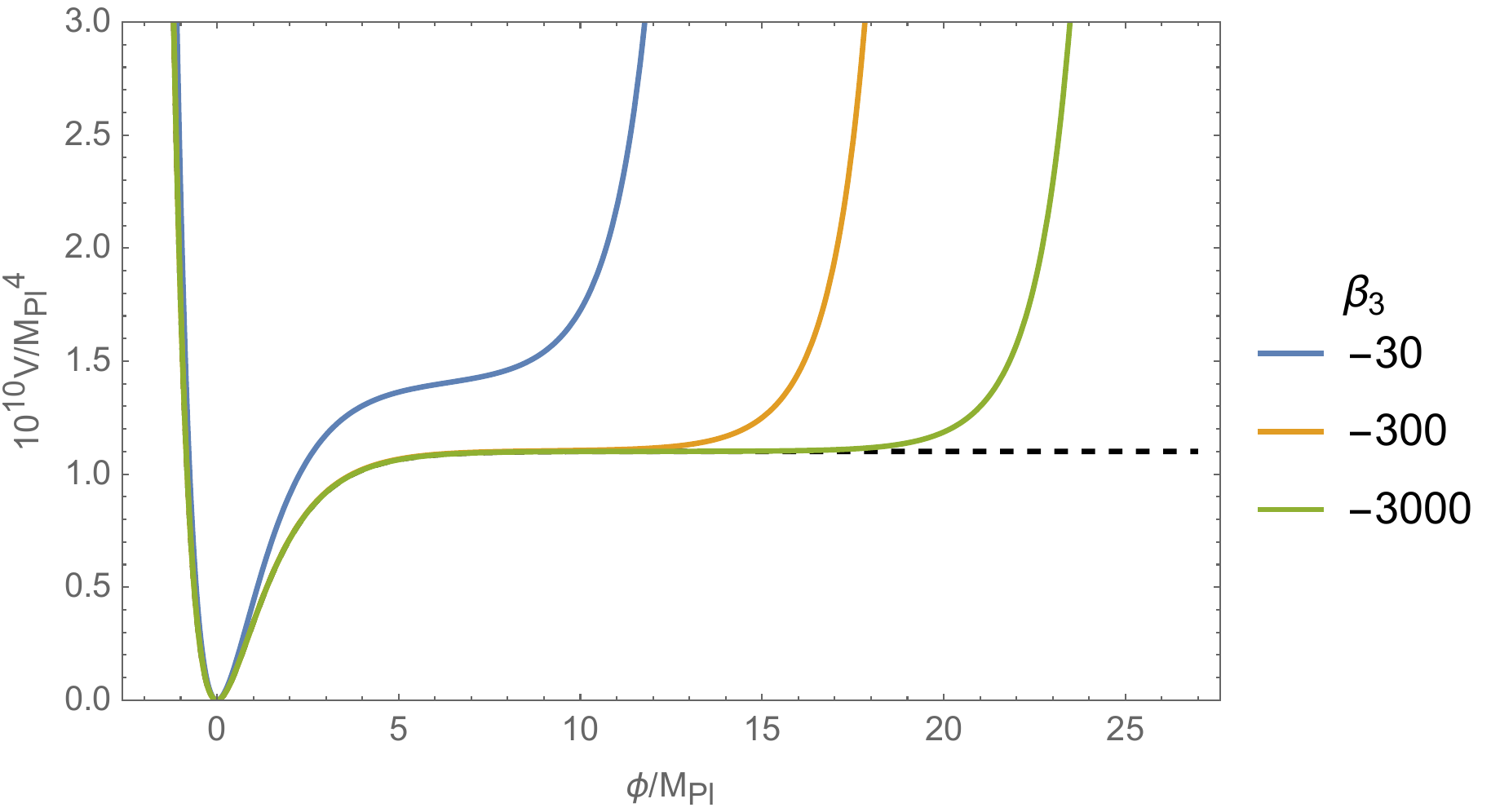}
  \caption{This is the potential~\eqref{eq-potential-example} with different parameter choices. 
  $ \alpha_3 =-2/3 $ and $ \alpha_R $ is determined by matching the scalar fluctuation $ \Delta_s^2 $ on CMB. 
  The black dashed line corresponds to the Starobinsky limit $ \beta_3 \to -\infty $. }
  \label{fig-pot}
\end{figure}
Shifting the minimum of the potential to the origin 
\begin{align}
    \phi_{\rm min} = \sqrt{\frac{2}{3}} \frac{\Mpl}{\alpha_3} \ln \left( -3\beta_3 +\sqrt{1+9\beta_3^2} \right) ~, 
\end{align}
we arrive at our final expression of the potential 
\begin{align}\label{eq-potential-example}
    V (\phi) = \frac{\Mpl^4}{144 \alpha_\text{R} \alpha_3^2} \left[ 3\beta_3 +\sinh \left( \sqrt{\frac{3}{2}}\alpha_3 \frac{\phi+\phi_{\rm min}}{\Mpl} \right) \right]^2 ~,
\end{align}
which is shown in Fig.~\ref{fig-pot}. 
One then can derive the potential slow-roll parameter as 
\begin{align}
    \epsilon_V &\equiv \frac{\Mpl^2}{2} \left( \frac{\partial V/\partial\varphi}{V} \right)^2 =3\alpha_3^2 \left[ \frac{\cosh \left( \sqrt{\frac{3}{2}} \alpha_3 \frac{\varphi-\varphi_0}{\Mpl} \right)}{ 3\beta_3 + \sinh \left( \sqrt{\frac{3}{2}} \alpha_3 \frac{\varphi-\varphi_0}{\Mpl} \right)} \right]^2 ~, \\
    \eta_V &\equiv \Mpl^2 \frac{\partial^2 V/\partial \varphi^2}{V} = 3 \alpha_3^2 \frac{\cosh \left( \sqrt{6} \alpha_3 \frac{\varphi-\varphi_0}{\Mpl} \right) + 3\beta_3 \sinh \left( \sqrt{\frac{3}{2}} \alpha_3 \frac{\varphi-\varphi_0}{\Mpl}\right) }{\left[ 3\beta_3 + \sinh \left( \sqrt{\frac{3}{2}} \alpha_3 \frac{\varphi-\varphi_0}{\Mpl} \right) \right]^2} ~,
\end{align}
where we changed the notation to match that in the previous section.
After expansion with respect to small $ \lambda $, we can map the parameters to Eqs.~\eqref{eq-approx-ns} and \eqref{eq-approx-r} by 
\begin{align}
    \gamma &= -\sqrt{\frac{2}{3}} \frac{1}{\alpha_3} ~,~
    \frac{V_0}{\lambda \Mpl V_1} = -\frac{3}{2} \beta_3 ~.
\end{align}
\begin{figure}[t]
  \centering
  \includegraphics[width=0.48\linewidth]{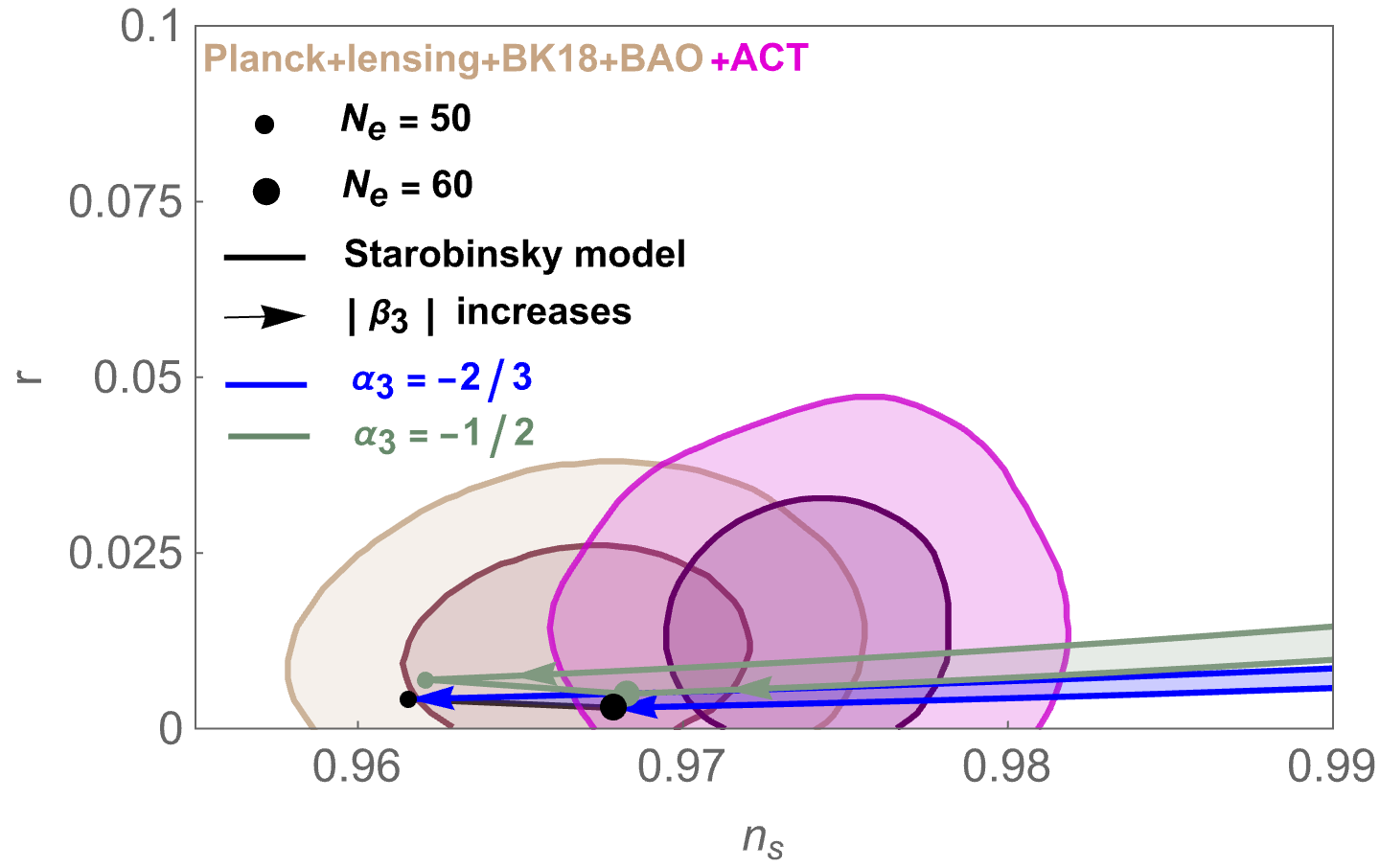}
  \includegraphics[width=0.48\linewidth]{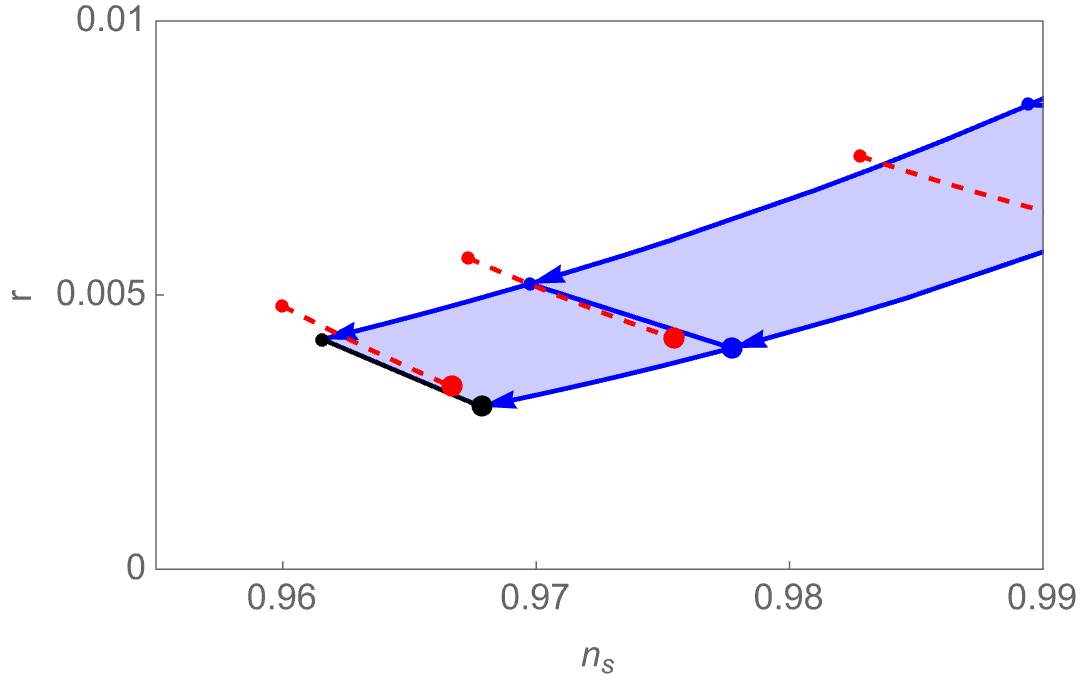}
  \caption{Predictions of spectral index $ n_s $ and tensor-to-scalar ratio $ r $ from the model given in Eq.~\eqref{eq-model} as an example of regularized pole inflation. 
  \textit{Left}: The constraint contours are directly taken from Fig. 5 in Ref.~\cite{BICEP:2021xfz} and Fig. 10 in Ref.~\cite{ACT:2025tim}.
  The left contours are constraints combining Planck~\cite{Planck:2018jri} with BICEP/Keck~\cite{BICEP:2021xfz} at pivot scale $ k/a_0=0.05 \,\mathrm{Mpc}^{-1} $ (and Ref.~\cite{BICEP:2021xfz} has assumed the tensor spectral index $ n_t =0 $), with $1 \sigma$ and $ 2\sigma $ regions respectively. 
  The right are those combined with ACT at pivot scale $ k/a_0=0.05 \,\mathrm{Mpc}^{-1} $. 
  In this example, $ \alpha_3 =-2/3 $ coincides with the predictions from the Starobinsky model when $ \beta_3 \to -\infty $ (practically we have taken $ |\beta_3|=3\times 10^4 $ in numerical calculation). 
  The deep blue trajectory is obtained with full numerical solution by fixing $ \alpha_3=-2/3 $ while changing $ \beta_3 $ from $ -13 $ ($ n_s $ is too large and lies outside the figure) to $ -3\times 10^4 $ (corresponding to small $ n_s $).
  Since large-$ |\beta_3| $ limit leads to the Starobinsky model, the predictions approach the black line as $ |\beta_3| $ increases. 
  The green trajectory is for $ \alpha_3 = -1/2 $ with the same range of $ \beta_3 $. 
  \textit{Right}: This is the zoom-in of the left panel and only the $ \alpha_3 =-2/3 $ trajectory is kept. 
  We also show the results calculated with the approximated formulae~\eqref{eq-approx-ns} and \eqref{eq-approx-r} in red dashed lines for $ \alpha_3 =-2/3 $ and different choices of $ \beta_3 $. 
  From the left to the right, the red dashed lines correspond to $ \beta_3 = -3\times 10^4 $, $ \beta_3 = -30 $, and $ \beta_3 = -17 $, 
  and the black and blue solid lines are calculated numerically by exact solutions with corresponding $ \beta_3 $'s. 
  One can see that the approximated formulae work well for the observationally relevant range of parameters, but generally they underestimate the prediction of $ n_s $, including that of the Starobinsky inflation limit.
  As an interesting consequence, in the Starobinsky model case, the predictions by the leading contributions in large $ N_e $ expansion is outside the favored regime by ACT, but the results from exact solutions can still reside in the observationally favored contour at the $2 \sigma$ level when $ N_e \simeq 60 $.} 
  \label{fig-nsr}
\end{figure}
In Fig.~\ref{fig-nsr}, we can compare the predictions from our example model~\eqref{eq-model} with CMB observation from the latest data released by ACT. 
One can indeed see that the regularization of pole can increase $ n_s $. 
It is worth pointing out that the leading order contributions in the large-$ N_e $ expansion generally underestimate $ n_s $ while the ACT result favors larger values compared with previous CMB observation, which leads to seemingly disfavor of the Starobinsky model even for $ N_e =60 $. 
However, the exact numerical result for $ N_e \simeq 60 $ in this model is still inside the ACT contour at the $2 \sigma$ level, as seen in the left panel of Fig.~\ref{fig-nsr}.

\section{Conclusion}\label{sec:conclusion}

In this work, we have proposed the regularized pole inflation, where the pole in the kinetic term of the inflaton field is regularized by a small parameter $ \lambda $.
As long as the $\lambda$ is small enough, the inflationary predictions are determined by the zero-th and first derivatives of the potential at the pole $\phi_0$, and hence the regularized pole inflation can still be regarded as an attractor model.
We have shown that, for such a small $\lambda$, the scalar spectral index $ n_s $ and the tensor-to-scalar ratio $ r $ increase with respect to the attractor predictions in proportional to $\lambda^2$.

Then, we have demonstrated that such a regularized pole inflation is naturally realized in the Einstein--Cartan gravity.
Allowing operators up to four dimensions, we have constructed a general Lagrangian in the Einstein--Cartan gravity.
Restricting ourselves to the case where the system does not yield $P(\phi,X)$ theory for simplicity, we have shown that the required structure of the kinetic function can be easily realized in this setup.
We have confirmed that the linear approximation of the potential around the regularized pole is sufficient to reproduce the exact calculations, which implies that the attractor based on the regularized pole is indeed valid to match the latest ACT results.

\section*{Acknowledgement}
M.\,He was supported by IBS under the project code, IBS-R018-D1.
M.\,Hong was supported by Grant-in-Aid for JSPS Fellows 23KJ0697.
K.\,M.\, was supported by JSPS KAKENHI Grant No.\ JP22K14044.

\bibliographystyle{utphys}
\bibliography{ref}

\providecommand{\href}[2]{#2}\begingroup\raggedright\begin{thebibliography}{10}

\bibitem{Starobinsky:1980te}
A.~A. Starobinsky, ``{A New Type of Isotropic Cosmological Models Without
  Singularity},'' \href{http://dx.doi.org/10.1016/0370-2693(80)90670-X}{{\em
  Phys. Lett. B} {\bfseries 91} (1980) 99--102}.

\bibitem{Sato:1980yn}
K.~Sato, ``{First Order Phase Transition of a Vacuum and Expansion of the
  Universe},'' {\em Mon. Not. Roy. Astron. Soc.} {\bfseries 195} (1981)
  467--479.

\bibitem{Guth:1980zm}
A.~H. Guth, ``{The Inflationary Universe: A Possible Solution to the Horizon
  and Flatness Problems},''
  \href{http://dx.doi.org/10.1103/PhysRevD.23.347}{{\em Phys. Rev. D}
  {\bfseries 23} (1981) 347--356}.

\bibitem{Mukhanov:1981xt}
V.~F. Mukhanov and G.~V. Chibisov, ``{Quantum Fluctuations and a Nonsingular
  Universe},'' {\em JETP Lett.} {\bfseries 33} (1981) 532--535.

\bibitem{Linde:1981mu}
A.~D. Linde, ``{A New Inflationary Universe Scenario: A Possible Solution of
  the Horizon, Flatness, Homogeneity, Isotropy and Primordial Monopole
  Problems},'' \href{http://dx.doi.org/10.1016/0370-2693(82)91219-9}{{\em Phys.
  Lett. B} {\bfseries 108} (1982) 389--393}.

\bibitem{Albrecht:1982wi}
A.~Albrecht and P.~J. Steinhardt, ``{Cosmology for Grand Unified Theories with
  Radiatively Induced Symmetry Breaking},''
  \href{http://dx.doi.org/10.1103/PhysRevLett.48.1220}{{\em Phys. Rev. Lett.}
  {\bfseries 48} (1982) 1220--1223}.

\bibitem{Sato:2015dga}
K.~Sato and J.~Yokoyama, ``{Inflationary cosmology: First 30+ years},''
  \href{http://dx.doi.org/10.1142/S0218271815300256}{{\em Int. J. Mod. Phys. D}
  {\bfseries 24} no.~11, (2015) 1530025}.

\bibitem{BICEP:2021xfz}
{\bfseries BICEP, Keck} Collaboration, P.~A.~R. Ade {\em et~al.}, ``{Improved
  Constraints on Primordial Gravitational Waves using Planck, WMAP, and
  BICEP/Keck Observations through the 2018 Observing Season},''
  \href{http://dx.doi.org/10.1103/PhysRevLett.127.151301}{{\em Phys. Rev.
  Lett.} {\bfseries 127} no.~15, (2021) 151301},
  \href{http://arxiv.org/abs/2110.00483}{{\ttfamily arXiv:2110.00483
  [astro-ph.CO]}}.

\bibitem{Cervantes-Cota:1995ehs}
J.~L. Cervantes-Cota and H.~Dehnen, ``{Induced gravity inflation in the
  standard model of particle physics},''
  \href{http://dx.doi.org/10.1016/0550-3213(95)00128-X}{{\em Nucl. Phys. B}
  {\bfseries 442} (1995) 391--412},
  \href{http://arxiv.org/abs/astro-ph/9505069}{{\ttfamily
  arXiv:astro-ph/9505069}}.

\bibitem{Bezrukov:2007ep}
F.~L. Bezrukov and M.~Shaposhnikov, ``{The Standard Model Higgs boson as the
  inflaton},'' \href{http://dx.doi.org/10.1016/j.physletb.2007.11.072}{{\em
  Phys. Lett. B} {\bfseries 659} (2008) 703--706},
  \href{http://arxiv.org/abs/0710.3755}{{\ttfamily arXiv:0710.3755 [hep-th]}}.

\bibitem{Barvinsky:2008ia}
A.~O. Barvinsky, A.~Y. Kamenshchik, and A.~A. Starobinsky, ``{Inflation
  scenario via the Standard Model Higgs boson and LHC},''
  \href{http://dx.doi.org/10.1088/1475-7516/2008/11/021}{{\em JCAP} {\bfseries
  11} (2008) 021}, \href{http://arxiv.org/abs/0809.2104}{{\ttfamily
  arXiv:0809.2104 [hep-ph]}}.

\bibitem{Kallosh:2013maa}
R.~Kallosh and A.~Linde, ``{Non-minimal Inflationary Attractors},''
  \href{http://dx.doi.org/10.1088/1475-7516/2013/10/033}{{\em JCAP} {\bfseries
  10} (2013) 033}, \href{http://arxiv.org/abs/1307.7938}{{\ttfamily
  arXiv:1307.7938 [hep-th]}}.

\bibitem{Kallosh:2013tua}
R.~Kallosh, A.~Linde, and D.~Roest, ``{Universal Attractor for Inflation at
  Strong Coupling},''
  \href{http://dx.doi.org/10.1103/PhysRevLett.112.011303}{{\em Phys. Rev.
  Lett.} {\bfseries 112} no.~1, (2014) 011303},
  \href{http://arxiv.org/abs/1310.3950}{{\ttfamily arXiv:1310.3950 [hep-th]}}.

\bibitem{Giudice:2014toa}
G.~F. Giudice and H.~M. Lee, ``{Starobinsky-like inflation from induced
  gravity},'' \href{http://dx.doi.org/10.1016/j.physletb.2014.04.020}{{\em
  Phys. Lett. B} {\bfseries 733} (2014) 58--62},
  \href{http://arxiv.org/abs/1402.2129}{{\ttfamily arXiv:1402.2129 [hep-ph]}}.

\bibitem{Pallis:2014dma}
C.~Pallis, ``{Induced-Gravity Inflation in no-Scale Supergravity and Beyond},''
  \href{http://dx.doi.org/10.1088/1475-7516/2014/08/057}{{\em JCAP} {\bfseries
  08} (2014) 057}, \href{http://arxiv.org/abs/1403.5486}{{\ttfamily
  arXiv:1403.5486 [hep-ph]}}.

\bibitem{Kallosh:2014laa}
R.~Kallosh, A.~Linde, and D.~Roest, ``{The double attractor behavior of induced
  inflation},'' \href{http://dx.doi.org/10.1007/JHEP09(2014)062}{{\em JHEP}
  {\bfseries 09} (2014) 062}, \href{http://arxiv.org/abs/1407.4471}{{\ttfamily
  arXiv:1407.4471 [hep-th]}}.

\bibitem{Mosk:2014cba}
B.~Mosk and J.~P. van~der Schaar, ``{Chaotic inflation limits for non-minimal
  models with a Starobinsky attractor},''
  \href{http://dx.doi.org/10.1088/1475-7516/2014/12/022}{{\em JCAP} {\bfseries
  12} (2014) 022}, \href{http://arxiv.org/abs/1407.4686}{{\ttfamily
  arXiv:1407.4686 [hep-th]}}.

\bibitem{Pieroni:2015cma}
M.~Pieroni, ``{$\beta$-function formalism for inflationary models with a non
  minimal coupling with gravity},''
  \href{http://dx.doi.org/10.1088/1475-7516/2016/02/012}{{\em JCAP} {\bfseries
  02} (2016) 012}, \href{http://arxiv.org/abs/1510.03691}{{\ttfamily
  arXiv:1510.03691 [hep-ph]}}.

\bibitem{Ellis:2013nxa}
J.~Ellis, D.~V. Nanopoulos, and K.~A. Olive, ``{Starobinsky-like Inflationary
  Models as Avatars of No-Scale Supergravity},''
  \href{http://dx.doi.org/10.1088/1475-7516/2013/10/009}{{\em JCAP} {\bfseries
  10} (2013) 009}, \href{http://arxiv.org/abs/1307.3537}{{\ttfamily
  arXiv:1307.3537 [hep-th]}}.

\bibitem{Ferrara:2013rsa}
S.~Ferrara, R.~Kallosh, A.~Linde, and M.~Porrati, ``{Minimal Supergravity
  Models of Inflation},''
  \href{http://dx.doi.org/10.1103/PhysRevD.88.085038}{{\em Phys. Rev. D}
  {\bfseries 88} no.~8, (2013) 085038},
  \href{http://arxiv.org/abs/1307.7696}{{\ttfamily arXiv:1307.7696 [hep-th]}}.

\bibitem{Kallosh:2013yoa}
R.~Kallosh, A.~Linde, and D.~Roest, ``{Superconformal Inflationary
  $\alpha$-Attractors},'' \href{http://dx.doi.org/10.1007/JHEP11(2013)198}{{\em
  JHEP} {\bfseries 11} (2013) 198},
  \href{http://arxiv.org/abs/1311.0472}{{\ttfamily arXiv:1311.0472 [hep-th]}}.

\bibitem{Kallosh:2014rga}
R.~Kallosh, A.~Linde, and D.~Roest, ``{Large field inflation and double
  $\alpha$-attractors},'' \href{http://dx.doi.org/10.1007/JHEP08(2014)052}{{\em
  JHEP} {\bfseries 08} (2014) 052},
  \href{http://arxiv.org/abs/1405.3646}{{\ttfamily arXiv:1405.3646 [hep-th]}}.

\bibitem{Carrasco:2015pla}
J.~J.~M. Carrasco, R.~Kallosh, and A.~Linde, ``{$\alpha $-Attractors: Planck,
  LHC and Dark Energy},'' \href{http://dx.doi.org/10.1007/JHEP10(2015)147}{{\em
  JHEP} {\bfseries 10} (2015) 147},
  \href{http://arxiv.org/abs/1506.01708}{{\ttfamily arXiv:1506.01708
  [hep-th]}}.

\bibitem{Roest:2015qya}
D.~Roest and M.~Scalisi, ``{Cosmological attractors from
  \ensuremath{\alpha}-scale supergravity},''
  \href{http://dx.doi.org/10.1103/PhysRevD.92.043525}{{\em Phys. Rev. D}
  {\bfseries 92} (2015) 043525},
  \href{http://arxiv.org/abs/1503.07909}{{\ttfamily arXiv:1503.07909
  [hep-th]}}.

\bibitem{Linde:2015uga}
A.~Linde, ``{Single-field $\alpha$-attractors},''
  \href{http://dx.doi.org/10.1088/1475-7516/2015/05/003}{{\em JCAP} {\bfseries
  05} (2015) 003}, \href{http://arxiv.org/abs/1504.00663}{{\ttfamily
  arXiv:1504.00663 [hep-th]}}.

\bibitem{Scalisi:2015qga}
M.~Scalisi, ``{Cosmological $\alpha$-attractors and de Sitter landscape},''
  \href{http://dx.doi.org/10.1007/JHEP12(2015)134}{{\em JHEP} {\bfseries 12}
  (2015) 134}, \href{http://arxiv.org/abs/1506.01368}{{\ttfamily
  arXiv:1506.01368 [hep-th]}}.

\bibitem{Galante:2014ifa}
M.~Galante, R.~Kallosh, A.~Linde, and D.~Roest, ``{Unity of Cosmological
  Inflation Attractors},''
  \href{http://dx.doi.org/10.1103/PhysRevLett.114.141302}{{\em Phys. Rev.
  Lett.} {\bfseries 114} no.~14, (2015) 141302},
  \href{http://arxiv.org/abs/1412.3797}{{\ttfamily arXiv:1412.3797 [hep-th]}}.

\bibitem{Broy:2015qna}
B.~J. Broy, M.~Galante, D.~Roest, and A.~Westphal, ``{Pole inflation
  \textemdash{} Shift symmetry and universal corrections},''
  \href{http://dx.doi.org/10.1007/JHEP12(2015)149}{{\em JHEP} {\bfseries 12}
  (2015) 149}, \href{http://arxiv.org/abs/1507.02277}{{\ttfamily
  arXiv:1507.02277 [hep-th]}}.

\bibitem{Terada:2016nqg}
T.~Terada, ``{Generalized Pole Inflation: Hilltop, Natural, and Chaotic
  Inflationary Attractors},''
  \href{http://dx.doi.org/10.1016/j.physletb.2016.07.058}{{\em Phys. Lett. B}
  {\bfseries 760} (2016) 674--680},
  \href{http://arxiv.org/abs/1602.07867}{{\ttfamily arXiv:1602.07867
  [hep-th]}}.

\bibitem{Fu:2022ypp}
C.~Fu and S.-J. Wang, ``{Primordial black holes and induced gravitational waves
  from double-pole inflation},''
  \href{http://dx.doi.org/10.1088/1475-7516/2023/06/012}{{\em JCAP} {\bfseries
  06} (2023) 012}, \href{http://arxiv.org/abs/2211.03523}{{\ttfamily
  arXiv:2211.03523 [astro-ph.CO]}}.

\bibitem{Aoki:2022bvj}
S.~Aoki, R.~Ishikawa, and S.~V. Ketov, ``{Pole inflation and primordial black
  holes formation in Starobinsky-like supergravity},''
  \href{http://dx.doi.org/10.1088/1361-6382/acb884}{{\em Class. Quant. Grav.}
  {\bfseries 40} no.~6, (2023) 065002},
  \href{http://arxiv.org/abs/2210.10348}{{\ttfamily arXiv:2210.10348
  [hep-th]}}.

\bibitem{Karamitsos:2021mtb}
S.~Karamitsos and A.~Strumia, ``{Pole inflation from non-minimal coupling to
  gravity},'' \href{http://dx.doi.org/10.1007/JHEP05(2022)016}{{\em JHEP}
  {\bfseries 05} (2022) 016}, \href{http://arxiv.org/abs/2109.10367}{{\ttfamily
  arXiv:2109.10367 [hep-th]}}.

\bibitem{Pallis:2021lwk}
C.~Pallis, ``{Pole-induced Higgs inflation with hyperbolic K\"ahler
  geometries},'' \href{http://dx.doi.org/10.1088/1475-7516/2021/05/043}{{\em
  JCAP} {\bfseries 05} (2021) 043},
  \href{http://arxiv.org/abs/2103.05534}{{\ttfamily arXiv:2103.05534
  [hep-ph]}}.

\bibitem{Dias:2018pgj}
M.~Dias, J.~Frazer, A.~Retolaza, M.~Scalisi, and A.~Westphal, ``{Pole
  N-flation},'' \href{http://dx.doi.org/10.1007/JHEP02(2019)120}{{\em JHEP}
  {\bfseries 02} (2019) 120}, \href{http://arxiv.org/abs/1805.02659}{{\ttfamily
  arXiv:1805.02659 [hep-th]}}.

\bibitem{Saikawa:2017wkg}
K.~Saikawa, M.~Yamaguchi, Y.~Yamashita, and D.~Yoshida, ``{Pole inflation in
  Jordan frame supergravity},''
  \href{http://dx.doi.org/10.1088/1475-7516/2018/01/031}{{\em JCAP} {\bfseries
  01} (2018) 031}, \href{http://arxiv.org/abs/1709.03440}{{\ttfamily
  arXiv:1709.03440 [hep-th]}}.

\bibitem{Kobayashi:2017qhk}
T.~Kobayashi, O.~Seto, and T.~H. Tatsuishi, ``{Toward pole inflation and
  attractors in supergravity : Chiral matter field inflation},''
  \href{http://dx.doi.org/10.1093/ptep/ptx166}{{\em PTEP} {\bfseries 2017}
  no.~12, (2017) 123B04}, \href{http://arxiv.org/abs/1703.09960}{{\ttfamily
  arXiv:1703.09960 [hep-th]}}.

\bibitem{Takahashi:2010ky}
F.~Takahashi, ``{Linear Inflation from Running Kinetic Term in Supergravity},''
  \href{http://dx.doi.org/10.1016/j.physletb.2010.08.029}{{\em Phys. Lett. B}
  {\bfseries 693} (2010) 140--143},
  \href{http://arxiv.org/abs/1006.2801}{{\ttfamily arXiv:1006.2801 [hep-ph]}}.

\bibitem{Nakayama:2010kt}
K.~Nakayama and F.~Takahashi, ``{Running Kinetic Inflation},''
  \href{http://dx.doi.org/10.1088/1475-7516/2010/11/009}{{\em JCAP} {\bfseries
  11} (2010) 009}, \href{http://arxiv.org/abs/1008.2956}{{\ttfamily
  arXiv:1008.2956 [hep-ph]}}.

\bibitem{ACT:2025fju}
{\bfseries ACT} Collaboration, T.~Louis {\em et~al.}, ``{The Atacama Cosmology
  Telescope: DR6 Power Spectra, Likelihoods and $\Lambda$CDM Parameters},''
  \href{http://arxiv.org/abs/2503.14452}{{\ttfamily arXiv:2503.14452
  [astro-ph.CO]}}.

\bibitem{ACT:2025tim}
{\bfseries ACT} Collaboration, E.~Calabrese {\em et~al.}, ``{The Atacama
  Cosmology Telescope: DR6 Constraints on Extended Cosmological Models},''
  \href{http://arxiv.org/abs/2503.14454}{{\ttfamily arXiv:2503.14454
  [astro-ph.CO]}}.

\bibitem{Dioguardi:2025mpp}
C.~Dioguardi and A.~Karam, ``{Palatini Linear Attractors Are Back in ACTion},''
  \href{http://arxiv.org/abs/2504.12937}{{\ttfamily arXiv:2504.12937 [gr-qc]}}.

\bibitem{Kim:2025dyi}
J.~Kim, X.~Wang, Y.-l. Zhang, and Z.~Ren, ``{Enhancement of primordial
  curvature perturbations in $R^3$-corrected Starobinsky-Higgs inflation},''
  \href{http://arxiv.org/abs/2504.12035}{{\ttfamily arXiv:2504.12035
  [astro-ph.CO]}}.

\bibitem{Antoniadis:2025pfa}
I.~Antoniadis, J.~Ellis, W.~Ke, D.~V. Nanopoulos, and K.~A. Olive, ``{How
  Accidental was Inflation?},''
  \href{http://arxiv.org/abs/2504.12283}{{\ttfamily arXiv:2504.12283
  [hep-ph]}}.

\bibitem{Salvio:2025izr}
A.~Salvio, ``{Independent connection in ACTion during inflation},''
  \href{http://arxiv.org/abs/2504.10488}{{\ttfamily arXiv:2504.10488
  [hep-ph]}}.

\bibitem{Gialamas:2025kef}
I.~D. Gialamas, A.~Karam, A.~Racioppi, and M.~Raidal, ``{Has ACT measured
  radiative corrections to the tree-level Higgs-like inflation?},''
  \href{http://arxiv.org/abs/2504.06002}{{\ttfamily arXiv:2504.06002
  [astro-ph.CO]}}.

\bibitem{Dioguardi:2025vci}
C.~Dioguardi, A.~J. Iovino, and A.~Racioppi, ``{Fractional attractors in light
  of the latest ACT observations},''
  \href{http://arxiv.org/abs/2504.02809}{{\ttfamily arXiv:2504.02809 [gr-qc]}}.

\bibitem{Berera:2025vsu}
A.~Berera, S.~Brahma, Z.~Qiu, R.~O.~Ramos, and G.~S. Rodrigues, ``{The early
  universe is $\textit{ACT}$-ing $\textit{warm}$},''
  \href{http://arxiv.org/abs/2504.02655}{{\ttfamily arXiv:2504.02655
  [hep-th]}}.

\bibitem{Aoki:2025wld}
S.~Aoki, H.~Otsuka, and R.~Yanagita, ``{Higgs-Modular Inflation},''
  \href{http://arxiv.org/abs/2504.01622}{{\ttfamily arXiv:2504.01622
  [hep-ph]}}.

\bibitem{Kallosh:2025rni}
R.~Kallosh, A.~Linde, and D.~Roest, ``{A simple scenario for the last ACT},''
  \href{http://arxiv.org/abs/2503.21030}{{\ttfamily arXiv:2503.21030
  [hep-th]}}.

\bibitem{Gao:2025onc}
Q.~Gao, Y.~Gong, Z.~Yi, and F.~Zhang, ``{Non-minimal coupling in light of
  ACT},'' \href{http://arxiv.org/abs/2504.15218}{{\ttfamily arXiv:2504.15218
  [astro-ph.CO]}}.

\bibitem{Brahma:2025dio}
S.~Brahma and J.~Calder\'on-Figueroa, ``{Is the CMB revealing signs of
  pre-inflationary physics?},''
  \href{http://arxiv.org/abs/2504.02746}{{\ttfamily arXiv:2504.02746
  [astro-ph.CO]}}.

\bibitem{Maeda:1987xf}
K.-i. Maeda, ``{Inflation as a Transient Attractor in R**2 Cosmology},''
  \href{http://dx.doi.org/10.1103/PhysRevD.37.858}{{\em Phys. Rev. D}
  {\bfseries 37} (1988) 858}.

\bibitem{Maeda:1988ab}
K.-i. Maeda, ``{Towards the Einstein-Hilbert Action via Conformal
  Transformation},'' \href{http://dx.doi.org/10.1103/PhysRevD.39.3159}{{\em
  Phys. Rev. D} {\bfseries 39} (1989) 3159}.

\bibitem{Trautman:2006fp}
A.~Trautman, ``{Einstein-Cartan theory},''
  \href{http://arxiv.org/abs/gr-qc/0606062}{{\ttfamily arXiv:gr-qc/0606062}}.

\bibitem{Racioppi:2024zva}
A.~Racioppi and A.~Salvio, ``{Natural metric-affine inflation},''
  \href{http://dx.doi.org/10.1088/1475-7516/2024/06/033}{{\em JCAP} {\bfseries
  06} (2024) 033}, \href{http://arxiv.org/abs/2403.18004}{{\ttfamily
  arXiv:2403.18004 [hep-ph]}}.

\bibitem{Karananas:2021zkl}
G.~K. Karananas, M.~Shaposhnikov, A.~Shkerin, and S.~Zell, ``{Matter matters in
  Einstein-Cartan gravity},''
  \href{http://dx.doi.org/10.1103/PhysRevD.104.064036}{{\em Phys. Rev. D}
  {\bfseries 104} no.~6, (2021) 064036},
  \href{http://arxiv.org/abs/2106.13811}{{\ttfamily arXiv:2106.13811
  [hep-th]}}.

\bibitem{Salvio:2022suk}
A.~Salvio, ``{Inflating and reheating the Universe with an independent affine
  connection},'' \href{http://dx.doi.org/10.1103/PhysRevD.106.103510}{{\em
  Phys. Rev. D} {\bfseries 106} no.~10, (2022) 103510},
  \href{http://arxiv.org/abs/2207.08830}{{\ttfamily arXiv:2207.08830
  [hep-ph]}}.

\bibitem{He:2024wqv}
M.~He, M.~Hong, and K.~Mukaida, ``{Starobinsky inflation and beyond in
  Einstein-Cartan gravity},''
  \href{http://dx.doi.org/10.1088/1475-7516/2024/05/107}{{\em JCAP} {\bfseries
  05} (2024) 107}, \href{http://arxiv.org/abs/2402.05358}{{\ttfamily
  arXiv:2402.05358 [gr-qc]}}.

\bibitem{He:unpublished}
M.~He, M.~Hong, and K.~Mukaida , In preparation.

\bibitem{Nieh:1981ww}
H.~T. Nieh and M.~L. Yan, ``{An Identity in Riemann-cartan Geometry},''
  \href{http://dx.doi.org/10.1063/1.525379}{{\em J. Math. Phys.} {\bfseries 23}
  (1982) 373}.

\bibitem{Nieh:2008btw}
H.~T. Nieh, \href{http://dx.doi.org/10.1142/9789812794185_0003}{``{A Torsional
  Topological Invariant},''} in {\em {Conference in Honor of C.N. Yang's 85th
  Birthday}: {Statistical Physics, High Energy, Condensed Matter and
  Mathematical Physics}}, pp.~29--37.
\newblock 2008.
\newblock \href{http://arxiv.org/abs/1309.0915}{{\ttfamily arXiv:1309.0915
  [gr-qc]}}.

\bibitem{Hojman:1980kv}
R.~Hojman, C.~Mukku, and W.~A. Sayed, ``{PARITY VIOLATION IN METRIC TORSION
  THEORIES OF GRAVITATION},''
  \href{http://dx.doi.org/10.1103/PhysRevD.22.1915}{{\em Phys. Rev. D}
  {\bfseries 22} (1980) 1915--1921}.

\bibitem{Nelson:1980ph}
P.~C. Nelson, ``{Gravity With Propagating Pseudoscalar Torsion},''
  \href{http://dx.doi.org/10.1016/0375-9601(80)90348-5}{{\em Phys. Lett. A}
  {\bfseries 79} (1980) 285}.

\bibitem{Castellani:1991et}
L.~Castellani, R.~D'Auria, and P.~Fre, {\em {Supergravity and superstrings: A
  Geometric perspective. Vol. 1: Mathematical foundations}}.
\newblock 1991.

\bibitem{Holst:1995pc}
S.~Holst, ``{Barbero's Hamiltonian derived from a generalized Hilbert-Palatini
  action},'' \href{http://dx.doi.org/10.1103/PhysRevD.53.5966}{{\em Phys. Rev.
  D} {\bfseries 53} (1996) 5966--5969},
  \href{http://arxiv.org/abs/gr-qc/9511026}{{\ttfamily arXiv:gr-qc/9511026}}.

\bibitem{Racioppi:2024pno}
A.~Racioppi, ``{$\tilde\xi$-attractors in metric-affine gravity},''
  \href{http://arxiv.org/abs/2411.08031}{{\ttfamily arXiv:2411.08031 [gr-qc]}}.

\bibitem{Gialamas:2024uar}
I.~D. Gialamas and A.~Racioppi, ``{Symmetry-breaking inflation in non-minimal
  metric-affine gravity},'' \href{http://arxiv.org/abs/2412.17738}{{\ttfamily
  arXiv:2412.17738 [gr-qc]}}.

\bibitem{Planck:2018jri}
{\bfseries Planck} Collaboration, Y.~Akrami {\em et~al.}, ``{Planck 2018
  results. X. Constraints on inflation},''
  \href{http://dx.doi.org/10.1051/0004-6361/201833887}{{\em Astron. Astrophys.}
  {\bfseries 641} (2020) A10},
  \href{http://arxiv.org/abs/1807.06211}{{\ttfamily arXiv:1807.06211
  [astro-ph.CO]}}.

\end{thebibliography}\endgroup

\end{document}